# Silicon heterojunction solar cells explored via noise spectroscopy: spatial selectivity and the influence of *a*-Si passivating layers

Kevin Davenport[1], Mark Hayward[1], C. T. Trinh[2], Klaus Lips[3], and Andrey Rogachev[1]

[1]Department of Physics and Astronomy, University of Utah, Salt Lake City, UT, USA.,
[2]Institute for Silicon Photovoltaics, Helmholtz-Zentrum Berlin für Materialien und Energie GmbH (HZB), Berlin, Germany
[3]Department ASPIN, Helmholtz-Zentrum Berlin für Materialien und Energie GmbH (HZB), Berlin, Germany

**Abstract**

We have employed state-of-the-art cross-correlation noise spectroscopy to study carrier dynamics in silicon heterojunction solar cells, complimented by SENTARUS simulations of the same devices. These cells were composed of a light absorbing *n*-doped crystalline silicon layer contacted by passivating layers of *i-a*-Si:H and doped *a*-Si:H electrode layers. The method provided a two-orders-of-magnitude improved sensitivity and allowed to resolution of three additional contributions to noise in addition to $1/f$ noise. We have observed shot noise with Fano factor close to unity. We have also observed a peculiar generation-recombination term, which presents only under light illumination with energy above 2 eV and thus reflects light absorption and carrier trapping in the *a*-Si:H layers. A second, low-frequency generation-recombination term was detected at temperatures below 100 K. We argue that it appears because the process of charge carrier transfer across *i-a*-Si:H occurs via an intermediate defect limited by tunneling above about 100 K and a thermally assisted process below this temperature. We also discuss the spatial selectivity of noise spectroscopy, namely the tendency of the method to amplify noise contributions from the most resistive element of the cell. Indeed, in our case, all three terms are linked to the passivating *i-a*-Si:H layer.

**Introduction**

Recent advances in the design and fabrication of silicon heterojunction solar cells (HJS) have boosted their operational parameters to record high values, exceeding 750 mV [1] for open-circuit voltage, $V_{OC}$, and 26 % efficiency [2]. Similar to conventional Si solar cells, crystalline silicon (c-Si) continues to be the main light-absorbing element in HJS. The charge-selective contacts are modified and made of *n*- and *p*-doped *hydrogenated amorphous silicon* (*a*-Si:H) which can be grown by the inexpensive low-temperature process of plasma-enhanced chemical vapor deposition, PECVD. Direct contact between c-Si and doped *a*-Si:H results in an interface with a high concentration of trap states [3,4] and leads to enhanced recombination which reduces the open circuit voltage of the cell. To mediate this problem, a very thin layer of intrinsic hydrogenated amorphous silicon (*i-a*-Si:H) is added on one or both sides of the c-Si wafer to passivate the interface [5,6,7].

The desired optimal performance of a silicon HJS imposes a series of strict requirements on the *a*-Si:H layer. First, it must be very thin so to avoid dissipating too much power or altering the absorption spectrum of the bulk. Second, it has to be grown and/or hydrogenated at a moderately high temperature with no or only moderate doping to have a low concentration of defects [6,8,9]. This temperature should not be too high as to promote epitaxial growth [10]. The realization of the importance of these (and many other) processes would not have been possible without the detailed structural and electronic characterization of the *a*-Si:H thin layers and interfaces. In some cases, very specialized methods such as near-UV photoelectron spectroscopy [3], surface photo-voltage measurements [4], and real-time spectroscopic ellipsometry [11] have been used. However, one might expect that due to chemical and physical interactions, the properties of a layer alter when it becomes part of the cell. Hence, further progress in the field, in particular reaching recently predicted 30% efficiency [12], requires development of complimentary specialized characterization tools that can identify and characterize the performance-limiting elements in fabricated solar cells despite their complex structure (multi-layers stack plus electrodes).

In this paper, we use current-noise spectroscopy to detect and characterize different electronic relaxation processes in a high-efficiency silicon HJSs. The well-known advantage of this method comes from resolution of the processes that occur at different time scales. The noise spectra reveal fingerprints of very specific electronic processes that conventional electrical characterization techniques are often not sensitive to. We argue here that current-noise spectroscopy also has a less obvious *spatial* selectivity, namely that it tends to magnify a contribution from the most resistive element of the stacked layers, in our case the passivating *i-a*-Si:H layers, and provides information on rate limiting electronic transitions such as recombination, tunneling, or thermal emission over a barrier.

Noise spectroscopy analyzes fluctuations of a signal from its equilibrium or steady-state value and is widely used for the characterization of defects and electronic relaxation processes in semiconducting devices [13,14,15]. It has been used in the past to study both doped and undoped *a*-Si:H [16], *a*-Si:H -based transistors [17], light-induced metastable changes in *a*-Si:H (Staebler-Wronski effect) [18], and more recently to evaluate defect states in crystalline solar cells [19]. In all these works, however, the focus has been exclusively on the low frequency $1/f$ contribution to noise.

The technical advantage of our work comes from employment of the current-noise cross-correlation technique [20], which provides two to four orders of magnitude improvement in the sensitivity and bandwidth of the measurements, giving access to regions of the noise spectrum which are typically hidden below the input noise floor. It is also very suitable for semiconductor devices with planar structure and high capacitance such as a typical photovoltaic cell. Using the technique, we were recently able to resolve the frequency-independent *shot* noise contribution in fluorescent [21] and multi-layered phosphorescent [22] organic light emitting diodes (OLEDs). In addition, analysis of the magnetic field dependence of the *generation-recombination* noise term [21] in fluorescent OLEDs helped us to identify the microscopic mechanism of the so-called organic magnetoresistance (OMAR) effect.

## Device Structure and Methods

In our experiments, we have studied solar cells with two slightly different structures, which we label SC1 and SC2; the SC1 solar cell is shown in Fig.1(a). The bulk of the device consists of a 145 μm-thick, n-doped (~$10^{15}$ dopants/cm$^3$), single crystal silicon layer prepared using the Czochralski method (*n*-Cz-Si layer). This layer is patterned with a 3-dimensional pyramidal texturing for increased light in-coupling by reducing external reflection [23]. Grown on either side of this bulk is a thin layer of hydrogenated intrinsic amorphous silicon (*i-a*-Si:H).

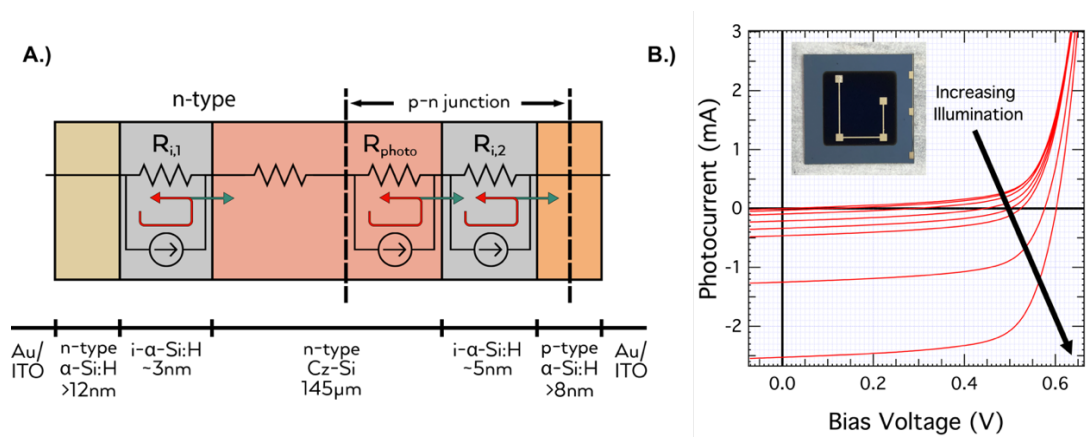

**Figure 1.** (A) Device structure diagram of the SC1-type HSC studied in this work including layer thicknesses. In the SC2 device, the 3nm-thick intrinsic amorphous layer is omitted. Overlaid is the proposed shot noise. (B) Typical *I(V)* characteristics under increasing illumination as indicated by the black arrow. The inset shows an image of the device under test; the dark blue active area is ~1cm$^2$.

The *p-n* heterojunction is formed with the addition of an ~8 nm thick layer of *p-a*-Si:H which serves as the emitter. This layer is doped to a degree that it can efficiently conduct current without incurring losses; in the devices studied, the emitter conductance is about $10^{-3}$ S/cm. On the reverse side of the device, a 12 nm thick layer of *n*-doped hydrogenated amorphous silicon (*n-a*-Si:H) is added that serves as a back surface field increasing the selectivity of the back contact and hence its collection ability for minority carriers [24].

Finally, both sides of the device are sealed with ITO, which acts as anti-reflection layer and back mirror in conjunction with the gold electrodes. An image of an actual test device studied is shown as an inset to Fig.1(b). The structure of SC2 solar cell is in all regards identical to the SC1 except that there is no *i-a*-Si:H layer between *n*-Cz-Si and *n-a*-Si:H.

Measurements of noise were carried out in a home-built, low-temperature flow cryostat in the temperature range 100-300 K and under illumination of light with varied wavelength. The cryostat and battery-powered LEDs were mounted on an optical table equipped with pneumatic vibration isolation. The details of the cross-correlation method are given in Refs. [20,25]. This method greatly decreases the noise of the front amplifiers; this reduction, however, is not complete with a theoretical noise floor of

$$S_{floor} = 2e_n^2 \left[ \frac{1}{R_D}\left(\frac{1}{R_D} + \frac{1}{R_f}\right) + \omega^2 C_D(C_D + C_i + C_{stray}) \right], \quad (1)$$

where $e_n$ is the input voltage noise of the operational amplifiers, $R_D$ and $C_D$ are the resistance and capacitance of the device under test respectively, $R_F$ is the feedback resistance of the transimpedance amplifiers, $C_i$ is the input capacitance of each channel, and $C_{stray}$ is the stray capacitance of the system. In our apparatus, the front transimpedance amplifiers are placed inside of the cryostat and operate low temperatures. The very short distance between the amplifiers and the sample (less than 5 cm) greatly reduces $C_{stray}$ and overall sensitivity to external noise sources. Using this system, we have been able to probe signals 4 orders of magnitude below the noise floor of the front amplifiers.

Notice that from Eq. 1, the *capacitance* of a device results in a term increasing with frequency as $f^2$. In principle, capacitance of multilayered devices can be a rather complex function of voltage [26]; it can also display useful correlations with noise (for example, in OLEDs we found one-to-one correlation between negative capacitance and generation-recombination noise term [27,28]). Unfortunately, our impedance measurements, not shown here, could resolve only a single contribution due to a capacitance of the *p-n* heterojunction, which displayed the expected divergent behavior $1/C^2 \propto (V_{OC} - V)$ observed in previous works [29,30]. Close to $V_{OC}$, divergent capacitance produces a very large $f^2$-term in the noise data (second term in Eq.1). Because this term obscures all other contributions, we have not studied noise dependence on bias and limited our measurements to the short-circuit configuration ($V = 0$) under different light intensities. The noise floor of our setup and high-frequency capacitive upturn have been approximated by data taken in the dark with 50,000 iterations. These contributions have subsequently been subtracted from the raw data obtained for different light intensities; the resulting noise curves are shown and analyzed in the rest of the paper.

## Experimental Results

In Fig. 2(a), we show the current noise power spectra versus frequency for the SC1 device under illumination by a red LED source ($\lambda = 625 \pm 5\ nm$) and at room temperature. Each curve corresponds to averaging over approximately 1000 iterations; the legend indicates the short-circuit current. Two clear features emerge from these spectra. First is a strong $1/f$ term extending to approximately 1 kHz. The second feature is the clear plateau of a frequency-independent term. The high-frequency features observed in the plateaus around 50 kHz are artifacts from an external source, likely injected into the system due to a ground loop.

To analyze the data, we fit each curve to following equation,

$$S_I = S_1 + \frac{S_2}{f^a} \quad (2)$$

Here, the first term represents the frequency-independent term comprised of shot noise and thermal noise. The second term models $1/f$-like flicker noise. Fig. 2(b) illustrates the importance of the cross-correlation method in detecting the true frequency-independent plateau. The dashed line represents the single channel signal, appearing an order of magnitude above the true signal.

From the fits, we find the exponent of the flicker component to $a \approx 1$ across all illumination intensities. In Fig. 2(c), we present the magnitude of the frequency-independent term, $S_1$, as a function of the short-circuit current, $J_{SC}$. This dependence is linear, giving strong evidence that this term represents shot noise present in the device. The fit to the equation $S_1 = 2eIF$, shown as a solid black line in Fig. 2(c), returns a reasonable value of the Fano factor $F \approx 0.9$. As we argue below, this indicates that the shot noise term represents a single process, most likely holes traversing the *i-a*-Si:H layers inside the *p-n* junction. It is important to notice that the correct value of the shot noise can only be obtained with cross-correlation method; for standard measurements, represented in our case by single channel output, the dashed line in Fig. 2(b), the frequency-independent term is heavily dominated by the noise from the front amplifier.

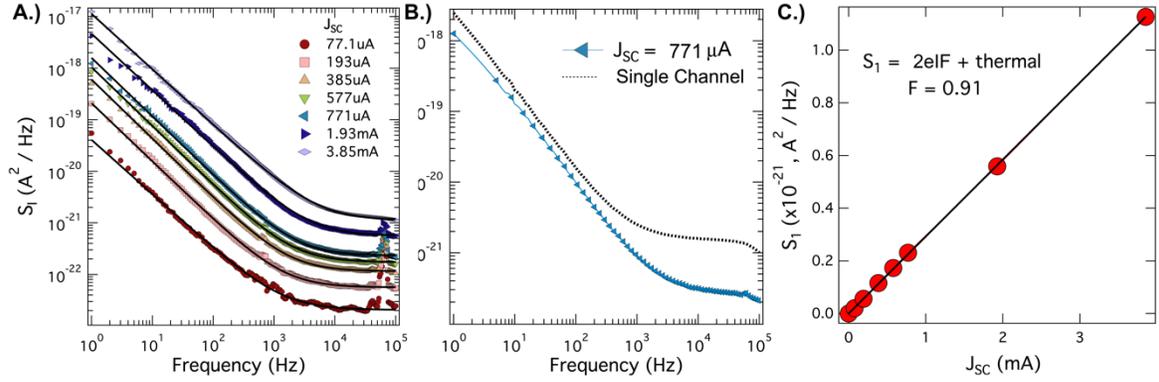

**Figure 2:** Current spectral density of the SHJ solar cell SC1 at short-circuit condition under red light illumination. (A) Frequency dependence of current noise power spectral density at 300K under illumination with varying intensity from a narrow-spectrum red LED ($\lambda = 625 \pm 5\, nm$). The legend indicates short-circuit DC current in the device. Black solid lines are fits to Eq. 1. (B) Comparison of a cross-correlated data set to that of a single channel illustrating the importance of the technique in detecting shot noise. (C) The magnitude of frequency-independent term extracted from fitting to Eq. 1 versus short-circuit current. Solid line is a linear fit to the data, corresponding to a Fano factor $F \approx 0.9$.

The noise spectra of the same device were also measured under illumination by a yellow LED source ($\lambda = 590 \pm 5\, nm$) and the power spectral densities are shown in Fig. 3(a). Compared to red light, a new feature clearly emerges between 1 kHz and 10 kHz; we name this term GR1. To account for this feature, we introduce a third term to the fitting equation,

$$S_I = S_1 + \frac{S_2}{f^a} + Re\left[\frac{S_3}{1 + (i\omega\tau)^{1-b}}\right] \quad (3)$$

This third term represents a generalization of generation-recombination noise, allowing for the dispersion of the relaxation time, $\tau$, as determined by the exponent $b$. Note that when $b = 0$, this term takes on the familiar form of a Lorentzian profile. To illustrate that this feature is not merely an artifact, Fig. 3(b) shows a representative data set for a single illumination with multiple fits to Eq. 3, each with a different term set to zero. The orange dashed line represents a fit identical to those in Fig. 2(a) in which $S_3 = 0$ making it obvious this feature was not present under red light. As is apparent, all three terms are needed to fit the data.

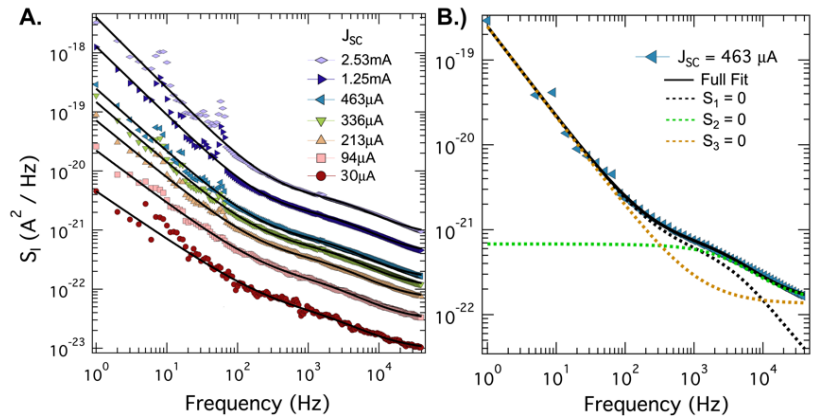

**Figure 3.** Current spectral density of the SHJ solar cell SC1 in short-circuit configuration under yellow light. (A) Frequency dependence of current noise power spectral density at 300 K under illumination with varying intensity from a narrow-spectrum yellow LED ($\lambda = 590 \pm 5\, nm$). The legend indicates short-circuit DC current in the device. Black solid lines are fits to Eq. 3. (B) A representative noise spectrum corresponding to $J_{SC} = 463\, \mu A$, where each term in Eq. 3 is sequentially set to zero to illustrate the presence of both frequency-independent and generation-recombination noise.

It is important to note that the frequency-independent term is also needed for a good fit; the black dashed line in Fig 3(b) shows the result when $S_1 = 0$. However, it was difficult for the curve fitting algorithm to capture the value of $S_1$ as a fit parameter. To account for this, we modeled it as shot noise $S_1 = 2eIF$ with a fixed Fano factor of $F = 0.9$ as found under red light.

The parameters extracted from these noise fits are presented in Fig. 4. To further verify the effect of light wavelength on noise spectra, we carried out several control measurements under illumination with blue light. In almost all details, the data were identical to that of yellow light. So, the emergence of the GR1 contribution has an energy threshold corresponding to the transition from red to yellow light and shows an electronic relaxation process that decreases with illumination level and has little dispersion ($b = 0.21$) and hence is difficult to be assigned to a process related to bandtails in the p-type a-Si:H selective contact layer. The typical relaxation time at room temperature of a few 10μs is expected for thermal emission over a barrier of about 0.45 eV as is present as valence-band offset between a-Si:H and c-Si [31].

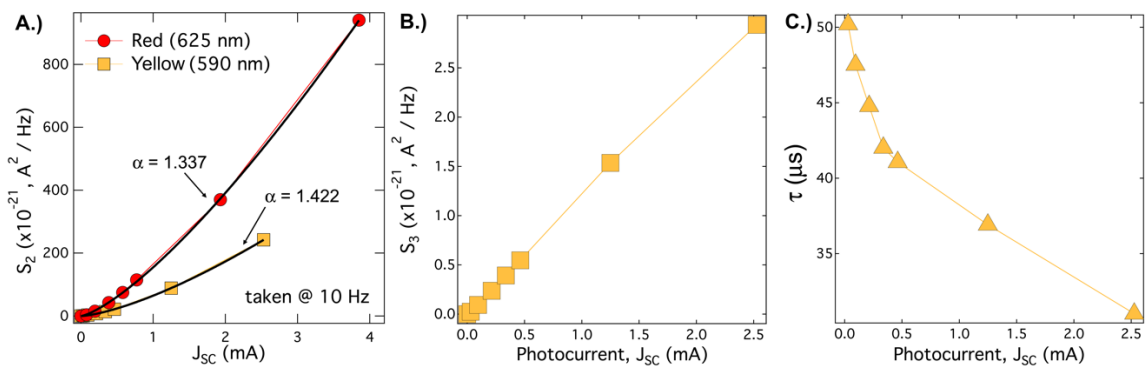

**Figure 4.** Extracted parameters from current noise spectra shown in Fig. 3. (A) The magnitude of the low frequency $1/f$ noise at 10 Hz as a function of short-circuit current for illumination with red and yellow light. The black curves represent fits to a power law, $S_2 = A \times J_{SC}^a$. (B) The magnitude of the generation-recombination term extracted from a fit to Eq. 3 as a function of $J_{SC}$; the value of the exponent of the generation-recombination term is $b = 0.21$. (C) The GR1 time constant plotted as a function of $J_{SC}$.

To clarify further the origin of the different noise terms, we performed the same series of tests on the SC2 device in which the passivating i-a-Si:H layer between the bulk n-Cz-Si and n-a-Si:H layer was omitted. In fact, this device was essential to our initial motivation to understand if noise spectroscopy can characterize trap states on the surface of crystalline silicon and the ability of *i-a*-Si:H to passivate them. In agreement with literature, the SC2 device exhibited a reduced open-circuit voltage by ~ 40mV at 1 sun. However, much to our surprise, the current noise spectra of this device under both red and yellow light illumination were found identical to what was observed in SC1 device. Despite the obvious effect on $V_{OC}$, the traps at the *n-a*-Si:H / *n*-Cz-Si interface do not appear to produce an observable additional noise signal.

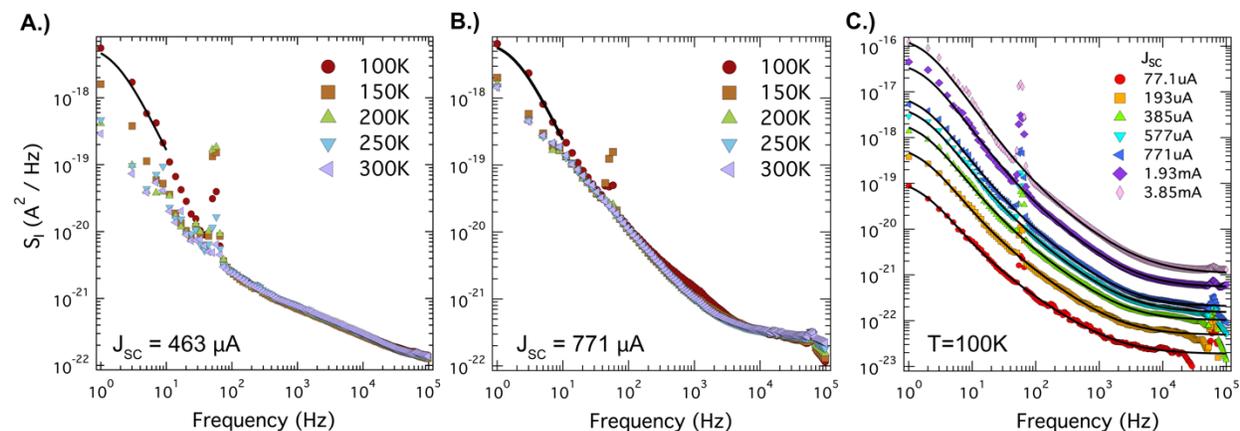

**Figure 5**. Current spectral density as a function of temperature of the SHJ solar cell SC1 at short-circuit condition. (A) The frequency and temperature dependence of the current noise spectra under yellow light illumination, illustrating the near temperature-independence of the GR1 noise term (frequency range 100 Hz – 10 kHz). This figure also illustrates an emerging low-temperature Lorentzian feature (GR2 term) with decreasing temperature. The black curves represent a fit to a standard Lorentzian line shape with $\tau \approx 0.1s$. The noise peak occurring at ∼ 80 Hz is caused by mechanical noise introduced by the flow of liquid nitrogen. (B) The frequency and temperature dependence of the current noise spectra under red light. Again, the GR2 spectrum with $\tau \approx 0.1s$ is detected below 100 K. (C) The full set of spectra under red light at 100 K. The black curves represent a fit to the sum of Eq. 3 and the single Lorentzian feature shown in the previous plots.

In our final experimental test, we performed noise measurements on the SC1 device at temperatures down to 100 K. The results are shown in Fig. 5; the sharp peak near 60 Hz is due to mechanical vibration introduced by the liquid nitrogen transfer line. The most noticeable change in spectra is the appearance of a new pronounced generation-recombination term at low frequencies and temperatures below about 100 K, which we will refer to as GR2. We found that this term can be approximated by a simple Lorentzian term, $S_4/[1+(f\tau)^2]$, with the single relaxation constant, $\tau \approx 0.1s$, being roughly the same both for red and yellow light. The total noise spectra, therefore, can be expressed by

$$S_I = S_1 + \frac{S_2}{f^a} + Re\left[\frac{S_3}{1+(i\omega\tau)^{1-b}}\right] + \frac{S_4}{[1+(f\tau)^2]} \qquad (4)$$

The magnitude of the shot noise and GR2 terms that result from the fits to Eq. 4 are shown in Fig. 6.

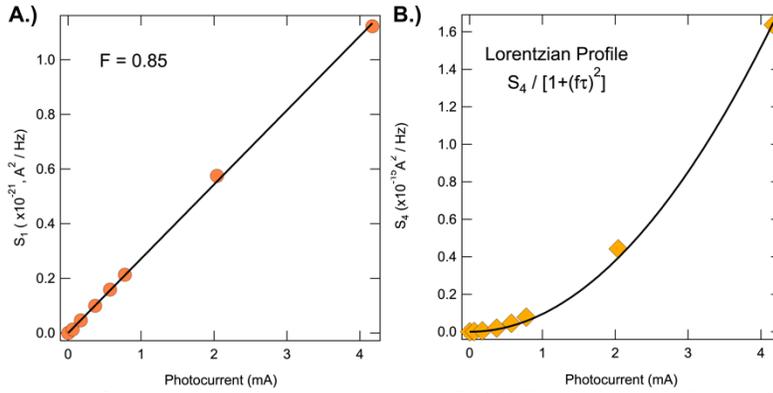

**Figure 6**. Extracted parameters from current noise spectra taken at T=100K and shown in Fig. 5. (A) The magnitude of the frequency-independent term extracted from the fits to eq. 4 in Fig. 5(c) as a function of photocurrent. The minimal reduction of the Fano factor as compared to the room-temperature data presented in Fig. 2 illustrates that this term is true shot noise. (B) The magnitude of the low-frequency Lorentzian term in eq. 4 seen to emerge below 150K as a function of photocurrent. The black line indicates a fit to an $J_{sc}^2$ dependence.

**Discussion**

The noise spectra of the silicon HJSs are unusually complex. We would like to start our analysis with the notion that in multi-layer devices, the most resistive element of the stack has the tendency to be a dominant source of observed noise. As a result, the method gains spatial selectivity; the downside is that the contribution from sources with low resistance become unobservable. To illustrate this property, we return to the noise circuit presented in Fig.1(a). To first approximation, the total noise in the device can be represented by a series of noise sources, $i_n$, each self-shorted by its own internal resistance, $R_n$. Here the sub-index *n* indicates distinct elements of the stack and the interfaces between them. From Kirchhoff's law, the total noise current seen at the contacts is given as

$$I_T = \sum_n \frac{i_n R_n}{R_T} \qquad (5)$$

where $R_T$ is the total device resistance. To be more specific, let us further assume that all generators $i_n$ represent uncorrelated shot noise terms with a full scale Shottky value $S_n = 2eI$. Thus, the total current noise seen at the electrodes is

$$S_T = \langle I_T^2 \rangle = \sum_n \langle i_n^2 \rangle \left(\frac{R_n}{R_T}\right)^2 = 2eI \sum_n \left(\frac{R_n}{R_T}\right)^2 = 2eIF, \qquad (6)$$

where the Fano factor, $F$, is determined by the resistor network and characterizes a suppression of the overall shot noise. For example, for $N$ identical sources connected in series, the expected Fano factor would be $F = 1/N$ [32]. We have verified the efficacy of this approach by performing noise measurements on identical forward-biased diodes, arranged in 1-, 2-, and 3-diode series configurations. As expected, the Fano factor scaled with the number of diodes, yielding $F = 0.48$ for two diodes and $F = 0.35$ for three.

Equations 5 and 6 describe a standard procedure for analyzing noise in electronic circuits composed of lumped elements. It has been successfully extended to analyze noise generated by hopping transport in inorganic [33] and organic semiconductors [21,22], where the high value of the Fano factor was taken as an indication of a small number of "hard" hops with a very high resistance. More recently, we have used this approach to analyze the transport processes which occur in a modern methylammonium lead triiodide perovskite solar cell [34].

It is important to note that the procedure is general and can be used when discussing *any* noise source. However, this extension is not straightforward as unlike shot and thermal noise, there is no fundamental mechanism fixing the magnitude and frequency dependence of $1/f$ and GR noise. Moreover, in real devices, the noise generator for an element of a stack, $i_n$, represents the incoherent sum of several independent noise sources. Fortunately, accounting for all possible terms is rarely needed in practice since for a particular frequency range, the noise generator is typically dominated by one or two microscopic processes.

Let us now discuss the individual contributions to the noise spectra of silicon HJSs as described in Eq. 4 and the possible microscopic processes generating them. This analysis was assisted by computer simulation using 1D and 2D numerical simulation package TCAD-SENTAURUS™ [35]; Figure 7 shows a sketch of the resulting band diagram. In the simulation, the $p$- and $n$-a-Si layers were modeled with a thickness of 20 nm and the $i$-a-Si layers with a thickness of 5 nm. Importantly, trap-assisted tunneling was enabled in both the doped and intrinsic films. The photogeneration of the vast majority of the carriers occurs far from the interfaces in the bulk (process 4) but photogeneration in the thin $a$-Si:H layer was also taken into account.

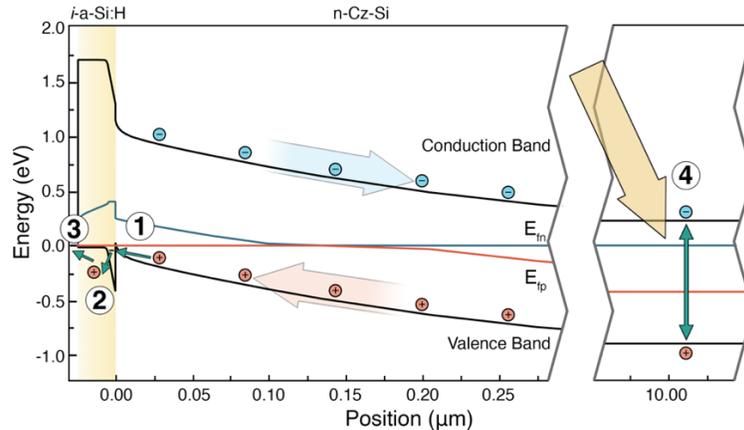

**Figure 7**. Band diagram sketch from TCAD-SENTAURUS™ simulation centered around the p$i$-a-Si:H / $i$-a-Si:H / n-Cz-Si interface; the n and p quasi Fermi levels are shown as blue and red dashed lines, respectively. The circled numbers highlight the noise processes contributing to the noise spectra: 1.) tunneling by holes into trap states inside the barrier at the interface, 2.) thermal emission from those trap states back into the conduction band, 3.) collection of holes at the front contact for recombination, and finally 4.) photogeneration of carriers.

**Frequency-independent term, $S_1$:** This term can unambiguously be assigned to shot noise as it follows the expected form $S_1 = 2eIF$ with almost the same value of Fano factor at room temperature, $F = 0.9$, and at 100 K, $F = 0.85$ as shown in Fig. 6(a). The lack of temperature dependence rules out thermal noise as an alternative mechanism behind this term.

In the SC1 HJS, there are potentially three processes producing shot noise and correspondently three sources connected in series, as illustrated in Fig 1(a). The first is a tunneling or thermally activated transition across the $n$-$a$-Si:H / $i$-$a$-Si:H / $n$-Cz-Si junction near the back of the device. The second is light-induced electron-hole pair generation in the $n$-Cz-Si bulk. Finally, the third is tunneling across the $n$-Cz-Si / $i$-$a$-Si:H / $p$-$a$-Si:H junction. Of the three, the third source has by far the highest resistance; it comes from $i$-$a$-Si:H layer, the higher resistance of the $p$-$a$-Si:H as compared to the *n-a*-Si:H, and the effect of the depletion layer [36]. Therefore, by Eqs. 5 and 6, the total measured shot noise of the HJS must represent the *n*-Cz-Si / *i-a*-Si:H / *p-a*-Si:H junction. This conclusion is supported by the fact that we see no difference between the Fano factors of the SC1 and SC2 devices.

Using the same argument, we notice that the near-unity value of the Fano factor strongly suggests that the transition across the *n*-Cz-Si / *i-a*-Si:H / *p-a*-Si:H junction occurs either in a single step or in a sequential multi-step propagation with one rate-limiting transition. In the language of Eqs. 5 and 6, this limiting transition has the highest "bottle-neck" resistance, although details of the microscopic process cannot be resolved using this technique alone.

**Flicker 1/*f*-like term, *S₂*:** Based on Eq.4, we again reason that this noise component comes from a process in the element of the HJS which has the highest resistance, the *n*-Cz-Si / *i-a*-Si:H / *p-a*-Si:H junction. Indirectly, this assertion is supported by the fact that the same magnitude of $1/f$ term is observed in both the SC1 and SC2 devices. The junction *n*-Cz-Si / *i-a*-Si:H / *p-a*-Si:H collects holes created in *n*-Cz-Si layer upon photoexcitation. The holes encounter a triangular potential barrier with a height $\Delta E_b \approx 0.4\ eV$ at the interface between *n*-Cz-Si and *i-a*-Si:H, as shown in Fig. 7. The barrier is composed of the *i-a*-Si:H (5 $nm$) and the depletion layer in *p-a*-Si:H, which is very thin ($\approx 5\ nm$) because of the high concentration of carriers [30]. Therefore, it is reasonable to assume that the origin of the flicker noise is the capture and release of holes as they tunnel to defect states within this barrier, labeled as processes 1 and 2 in Fig. 7.

The observed deviation in the dependence of the power spectral density on current, $S_2 = AJ_{SC}^\alpha$, (see Fig. 4a) from its standard value $\alpha \approx 2$ implies that the noise comes not from the existing defects but rather from ones induced by the light or current [37]. Two possible candidates are light induced migration of hydrogen atoms, which can modify local probability of tunneling across *i-a*-Si:H layer and trapped charges at interfaces, which can modify local height of the potential barriers.

**GR1 Generation-recombination term, *S₃*:** This term, shown in Fig. 5(a,b), has several peculiar properties. Its magnitude and time constant do not depend on temperature. Further, the magnitude does not follow the $J^2$ depenance (Fig 4b) expected for systems where charge carriers probe existing traps of defect states. The relaxation time, $30 - 50\mu s$ (Fig 4c), only weakly decreases (by a factor of two-fold) when electrical current changes by two orders of magnitude. Most remarkably, the GR1 term is absent under red light, when energy of photons is 1.98 eV, and appears under yellow light, when photons have only slightly larger energy of 2.10 eV. This term is also present under blue light with essentially with the same characteristics as shown here for yellow.

The only parameter in the silicon HJS that could match this energy threshold is the optical gap in amorphous silicon. The typically listed value for the gap is smaller, around 1.8 eV, however it has been experimentally observed that it grows with hydrogen concentration, reaching a value of 2.0 eV at 25% [38]. It has also been observed that the gap increases with the decreasing thickness of *a*-Si:H films, reaching 2.0 eV in films with $t < 3.5\ nm$ [39]. Therefore, we suggest that GR1 is likely due to a process initiated by the absorption of light in the amorphous silicon layer. This supposition is supported by a notion that the *a*-Si:H layer has very high resistance and hence its noise is greatly enhanced compared to other elements of the stack. The microscopic origin of this process is not understood, though the lack of temperature dependence suggests that it is likely a tunneling process.

Overall, the detection of this peculiar GR1 process suggests that the relaxation processes in these solar cells have unanticipated channels and that further study of the processes associated with *a*-Si:H and the *a*-Si:H /Cz-Si interface may lead to improvement in solar cell performance.

**GR2 Generation-recombination term, *S₄*:** The generic case of noise caused by the fluctuations in carrier density due to capture and release of carriers by identical trap levels is described by the equation [40]

$$S_{EC}(f) = \frac{S_4}{1+(f\tau)^2} = 2\pi \frac{4N_t}{Vn_0^2} \frac{\tau B(1-B)}{1+(f\tau)^2} I^2, \qquad (7)$$

where $N_t$ is the concentration of the traps, $V$ the volume of the sample, $B$ the filling factor of the trap site such that $0 \leq B \leq 1$, $n_0$ the carrier concentration, and $\tau = \tau_e\tau_c/(\tau_e + \tau_c)$ the combined relaxation time where $\tau_c$ and $\tau_e$ are mean capture and emission times, respectively. Equation 7 captures two characteristics of the GR2 process. First is a single relaxation time with near-zero dispersion, meaning it is created by an almost mono-energetic trap level. Second is $I^2$ dependence, which means that existing traps are probed. Also, from our generic arguments based on Eqs. 5 and 6, we expect that the relevant process takes place in the *n*-Cz-Si / *i-a*-Si:H / *p-a*-Si:H junction.

The most likely explanation for the emergence of this feature is the switch of the rate dependence of the emission of the holes through the triangular barrier from tunneling to thermal emission, process 2 in Fig. 7. At room temperature, the time scale of the tunneling step, process 1 in Fig. 7, is on the order of tens of microseconds whereas the time scale of processes 2 and 3 are in the picosecond range. With decreasing temperature, the timescale of thermal emission from trap states in the barrier increases to the point where it overtakes tunneling as the rate limiting process. The expected timescale of this process is set by the escape time from the trap states which can be estimated as $\tau \approx 1/\left[\upsilon \times \exp\left(-\frac{\Delta E}{k_b T}\right)\right]$, where $\upsilon \sim 10^{12}\ s^{-1}$ is the typical attempt frequency for silicon. The appearance of the GR2 term with $\tau \approx 100$ ms at temperature about 100 K places the estimate for the trap level at $\Delta E = 0.2 - 0.25$ eV.

## Conclusions

In this paper, we have studied the noise properties of silicon heterojunction solar cells in the temperature range 100-300 K and under illumination by red and yellow light. These measurements were done in the short-circuit configuration using a cross-correlation method in a current-monitoring configuration. We observe the noise spectra to display rich information content. First, we observe a shot-noise contribution with Fano factor close to 1. Second, the $1/f$ contribution was found to be temperature-independent and related to photocurrent as $J^\alpha$. The observed exponent $\alpha \approx 1.3 - 1.5$, instead of standard $\alpha = 2$, suggests that this contribution reflects not the existing defects but rather those induced by light and/or current. Light-induced migration of hydrogen atoms and/or light-induced charge states are possible candidates for this process. Third, we observe a generation-recombination term (GR1) that is only present under illumination by the light with energy above 2 eV, thus reflecting excitations in amorphous Si layers. Finally, a second generation-recombination term (GR2) with a single relaxation time appears at temperatures below around 100 K. The first of these GR terms results from the tunneling of holes through the bulk/$a$-Si:H interface and is observed to have a timescale ranging from 50 µs to 30 µs over the range of illuminations studied. The second emerges at the rate-limiting step at this interface switches from tunneling to the thermal emission out of the bandtail trap states.

We have also presented arguments that in multi-layer devices, the current noise spectra are typically dominated by the contributions from most resistive element in the device stack. As in the studied solar cells the most resistive elements are the passivating layers of intrinsic amorphous silicon, $i$-$a$-Si:H, we conclude that they are responsible for the shot noise, GR1 noise and $1/f$ noise terms.

The complex planar stacks are widespread in modern solar cells and light emitting diodes, so the unappreciated space selectivity of current noise spectroscopy could potentially make it a useful and non-invasive complimentary characterization tool, particularly when implemented in a cross-correlation configuration.

**Acknowledgment**
We acknowledge the preparation of the two silicon HJSs by Holger Rhein and Anna B. Morales-Vilches of PVcomB at HZB. KL is indebted to the Deutsche Forschungsgemeinschaft (DFG), which supported the research visits at the University of Utah through the priority program SPP1601. KD and AR gratefully acknowledge the support by NSF grants DMR1611421 and DMR1904221.

**Competing Interests**
The authors declare no competing interests related to this work.

**Data Availability Statement**
The datasets generated and analyzed in this work are available from the corresponding author upon reasonable request.